\documentclass[12pt,journal,compsoc]{article}

\usepackage{amssymb}
\usepackage{graphicx}
\usepackage{caption}
\usepackage{subcaption}
\usepackage{url}
\usepackage{wrapfig}
\usepackage{placeins}

\begin{document}


\title{Leveraging Semantic Web Technologies for Managing Resources in a Multi-Domain Infrastructure-as-a-Service Environment\footnote{This research is supported by NSF grants ACI-1032573, ACI-1245926 and DOE award ASCR DE-SC0005286.}}


%
%
\author{Yufeng Xin, Ilya Baldin \\\{yxin, ibaldin\}@renci.org\\RENCI/UNC Chapel Hill \\ \and Jeff Chase \\chase@cs.duke.edu\\ Duke University \\ \and  and Kemafor  Anyanwu \\ kogan@ncsu.edu\\NCSU}
%





%
%


\maketitle

\begin{abstract}
This paper reports on experience with using semantically-enabled network resource models to construct an operational multi-domain networked infrastructure-as-a-service (NIaaS) testbed called ExoGENI, recently funded through NSF's GENI project.  A defining property of NIaaS is the deep integration of network provisioning functions alongside the more common storage and computation provisioning functions. Resource provider topologies and user requests can be described using network resource models with common base classes for fundamental cyber-resources (links, nodes, interfaces) specialized via virtualization and adaptations between networking layers to specific technologies.

This problem space gives rise to a number of application areas where semantic web technologies become highly useful - common information models and resource class hierarchies simplify resource descriptions from multiple providers, pathfinding and topology embedding algorithms rely on query abstractions as building blocks. 

The paper describes how the semantic resource description models enable ExoGENI to autonomously instantiate on-demand virtual topologies of virtual machines provisioned from cloud providers and are linked by on-demand virtual connections acquired from multiple autonomous network providers to serve a variety of applications ranging from distributed system experiments to high-performance computing. 

\end{abstract}

\section{Introduction}
Cloud provider services like Amazon EC2, Microsoft Azure and RackSpace are examples of IaaS (Infrastructure-as-a-Service) public cloud providers. Modern open source technologies like OpenStack~\cite{OpenStack:www} and Eucalyptus~\cite{eucalyptus} permit the creation of private institutional IaaS clouds. In either case, through the use of a well-defined API, the properly authorized consumer can provision compute and storage resources for themselves. The virtual compute and storage infrastructure they get behaves similar to real infrastructure and is accessed remotely over commodity Internet. The networking resources within cloud provider infrastructure are provisioned implicitly. 

\begin{wrapfigure}{r}{0.5\textwidth}
\vspace{-0.3in}
  \begin{center}
    \includegraphics[width=0.48\textwidth]{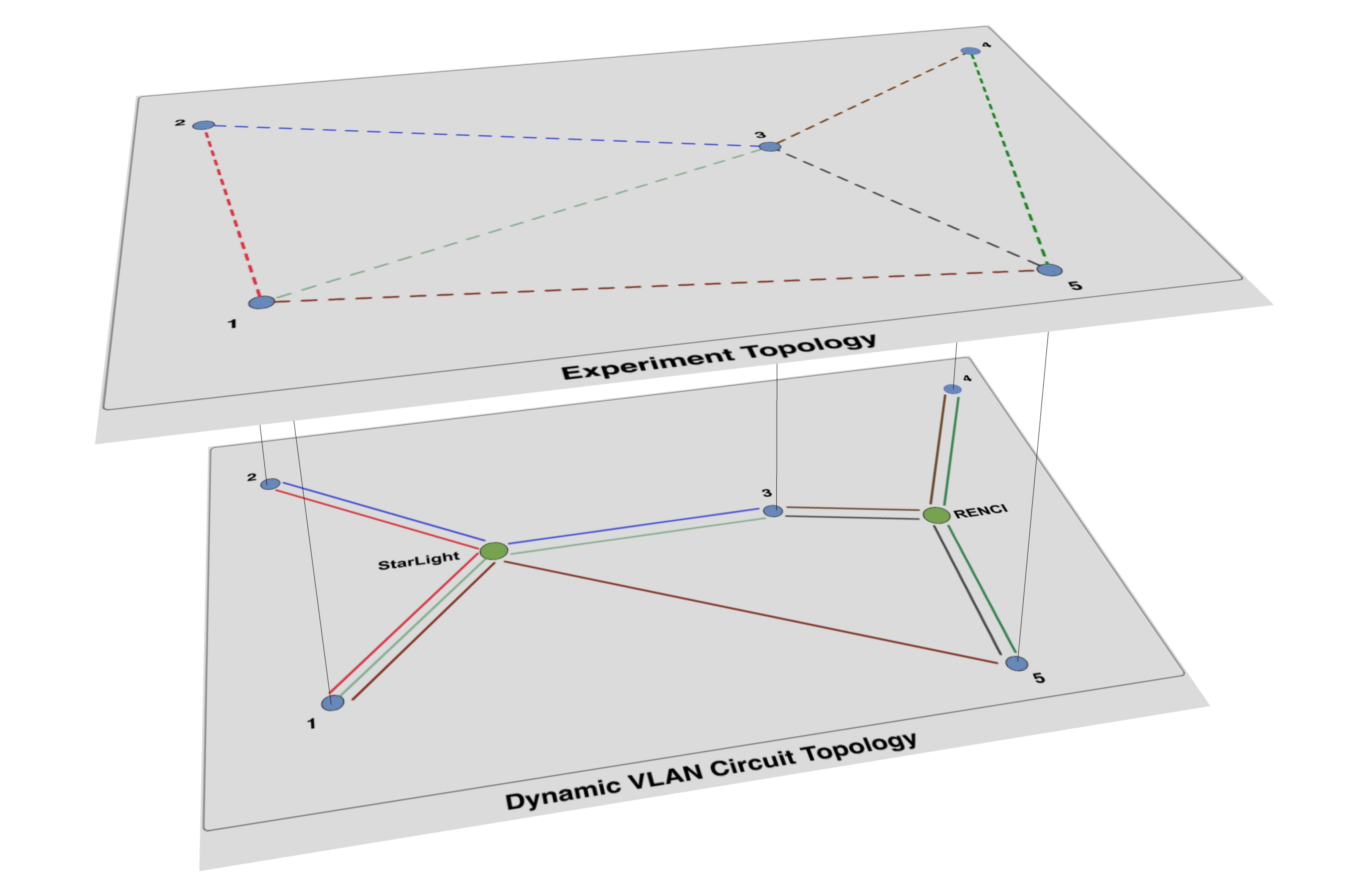}
  \end{center}
  \caption{Customer slice (top) assembled from multiple institutional cloud and network providers (bottom). }
  \vspace{-0.2in}
  \label{fig:slice}
\end{wrapfigure}

A less known type of IaaS providers are on-demand network services (in academic environments Internet2~\cite{i2:dcn}) that permit explicit virtualization of their resources by users {\em on-demand} - the creation of links between various points within their networks with well defined Quality of Service (QoS) characteristics like bandwidth, latency and jitter. Technologies used for this are typically VLANs or MPLS, although this is not relevant for the further discussions in this paper. What is critical is that for institutional clouds connected to these networks this enables a markedly different on-demand approach to building interconnects between them, distinct from the more common public cloud approach, where either permanently dedicated network connections or the best-effort commodity Internet is used as the interconnect. The type of performance isolation provided by this dynamic capability is required by many distributed experimental and production applications - the driving motivation behind our efforts.

We call our approach {\em Networked} Infrastructure-as-a-Service (NIaaS) because of the deep integration of network provisioning functions with the computational and storage provisioning functions. Our high-level goal is the enabling of a federation of multiple diverse resource providers, i.e. computational and storage institutional clouds, on-demand networks 
for the purpose of customer-driven on-demand creation of complex connected arrangements of compute, storage and network resources collected from those providers. Those arrangements, called '{\em slices}' are virtual network topologies of compute, storage and network resources as defined by the consumer. Slices are provisioned on-demand and persist for the duration of the consumer need. They serve as platforms for running multiple concurrent complex distributed computational activities that are isolated from one another. This contrasts with the grid approach of running multiple concurrent activities on shared infrastructure.

In federating multiple providers for this purpose a key problem is in {\em scheduling and orchestration of resource provisioning actions across  many cloud and network providers}, so resulting slice topologies mimic that of the customer request (see Figure~\ref{fig:slice}) in which semantic web technologies play a critical role. There are several motivating factors that make them particularly applicable to this environment. 

First, the orchestration process is heavily dependent on declarative descriptions of compute, storage and network resources in order to perform its work: the consumer must be able describe the desired slice topology, the resource provider must be able to describe resources available for orchestration and the system must maintain the state of the currently utilized resources. Semantic descriptions with their complex hierarchies of entity classes and property relationships and standardized vocabularies act as the common abstraction layer to which all other representations can be converted. Critically, these can be extended by individual providers to define classes and properties specific to their environment. The RDFS and OWL entailments allow common resource management and topology embedding algorithms to operate on the shared common classes, thus improving their portability. Considering the main goal of our work of enabling a multi-provider heterogeneous and {\em federated} NIaaS environment, having such a common and extendable way of describing resources is a critical property.


Second, common networking computation tasks, like path computation and virtual topology mapping, can be modeled as subgraph extractions on the semantic graph~\cite{detwiler2008regular}, that we discuss later in the paper. This allows new resource management algorithms to be built as procedural code heavily leveraging common operations abstracted as standardized queries that are independent of the programming environment and implemented efficiently in common toolsets. Using queries is motivated by similar goals as the development of database management systems to replace hardcoded file processing algorithms: i.e. enabling reuse and automatic optimization.

Third, rule engines can be used to perform additional processing on the models in a declarative, rather than procedural fashion, which makes them more portable and verifiable - a critical feature in complex distributed systems. Finally, once the representations are converted to semantic web formats (in our case OWL DL), they can be operated on using a large selection of mature tools used for querying (SPARQL) and inference (Pellet, Hermit~\cite{parsia2004pellet,shearer2008hermit}). 

The alternatives, as used today in many systems~\cite{PerfSonar,rspec,ec2}, are JSON or XML-based schemas, encoding only the syntax rules, making them hard to validate semantically. The relationships between object classes and roles are modeled as ad-hoc procedural code which differs from implementation to implementation, rather than explicit object and relationship hierarchy rigorously encoded in OWL. Resource management code operating on such representations lacks the ability to leverage common abstractions and optimizations. 

Our main contribution lies in designing a set of ontologies that are relevant to NIaaS problem space and constructing a production NIaaS system that actively uses semantic technologies for autonomous provisioning and managing such diverse resources at scale (note that in this paper we use the term `resource' when referring to computational, storage and network resources, rather than RDF resources). This system is called ExoGENI~\cite{exogeni-tridentcom12}, part of NSF-funded GENI (Global Environments for Network Innovations)  federation of testbeds supporting distributed large-scale experiments in computational and network sciences.

The following sections discuss related work and detail some of the uses of these technologies within ExoGENI.

\vspace{-0.1in}
\section{Related work}
{\bf Semantic descriptions of networks}

The initial building block for our work is an RDFS ontology called NDL ( {\em Network Description Language})~\cite{NDL:main,NDL:Overview,NDL:optical-nets} developed by network researchers at the University of Amsterdam. 

NDL is based on the ITU-T G.805 standard~\cite{ITU:G805},  {\it Generic functional architecture of transport networks} and provides an abstract informational model for connection-oriented transport networks. Transport networks carry multiple types of traffic, including Internet traffic, however unlike the Internet, they provide capabilities for provisioning bandwidth-on-demand in the form of channels at potentially different layers (optical, ethernet and so on). Connections at different layers within transport networks have server-client relationships with a server layer connection serving as an envelope for several client connections. As an example, multiple Ethernet VLANs (virtual links) can be carried inside a single optical wavelength. Critically, certain types of networking equipment are capable of {\em adaptations} from one layer to another, i.e. accepting one or more client connections (e.g. VLANs) and multiplexing them  onto a server connection at a lower layer (e.g. optical wavelength or timeslot). These capabilities act as constraints on pathfinding operations needed by topology embedding algorithms.
The primary use of NDL has been in GLIF~\cite{NDL:GLIF}, where it is used by individual network providers for sharing the details of their topologies with each other. 

We redefined NDL as an OWL DL ontology and called our variant {\em NDL-OWL}. The reason for redefining it was two-fold - it was a means of future-proofing our work, as we are interested in exploring the use of OWL DL constraints and inferences to assist in provisioning tasks and also OWL DL tools provide a richer set of capabilities compared to RDF/RDFS. 

{\bf Semantic descriptions in cloud technologies}


In mOSAIC~\cite{Moscato2011} the authors present a compute ontology based on a collection of cloud taxonomies (NIST~\cite{nist-cloud}, OCCI~\cite{Nyren2011}). This ontology is part of a larger effort to create a unified cloud API that is semantically enriched using elements of the ontology. The effort is concentrated on unifying the views of different cloud providers of varying types (SaaS, PaaS, IaaS) under a single API. Our own compute ontology is also loosely based on NIST and other taxonomies, but is focused only on a single provider type - IaaS, however is much richer in terms of its ability to describe network topologies. 

In~\cite{Haase} the authors present a system for enterprise cloud management that automatically catalogs resources within the enterprise from its various elements (compute resources, storage resources) and presents the accumulated semantic database for presentation via UI or analytics using SPARQL queries. The authors also describe a compute ontology focused on detailed infrastructure element descriptions. While the authors allude to other elements of their system that are capable of performing infrastructure provisioning tasks, their linkage to the semantically-enabled portion is not described. One interesting property of their system not currently present in ExoGENI is the ability to automatically collect and convert resource data into semantic triples. In ExoGENI, site operators must at the moment manually create semantic descriptions of their resources and their topologies using tools like Proteg\'{e}. 

In~\cite{Tahamtan2012} the authors describe a semantically-enabled system meant to assist cloud users to select the appropriate cloud provider based on a variety of requirements, from the underlying hosting hardware, to the availability of higher level business software and even their power use. As a whole, the system covers multiple XaaS cloud types (IaaS, PaaS, SaaS) and their ontology is cloud-consumer-centric. They use a semantic reasoner to derive additional facts based on input data, which can be used to aid user selection of the appropriate provider.

Semantic grid~\cite{Corcho2006a} effort is similar to our own, in that it tried to bring order to the representation of resources among grid providers, however they used different resource abstractions, grounded in the services approach i.e. not what a resource is (a node with this much CPU power, memory, disk), but what a service installed on the node does. The problems that they solve using semantic grid, like service composition, are different than the problems we deal with, namely topology embedding, because they use different types of constraints: which services can compose with which, rather than e.g. finding paths across a multidomain provider infrastructure across multiple adaptations.


\vspace{-0.1in}
\section{Using semantic models in ExoGENI}
\label{sec:using}

Our NIaaS system called ExoGENI~\cite{exogeni-tridentcom12,ExoGENI:www} serves as a production service for distributed experimental activities by computer scientists from multiple universities and labs. The testbed is funded by US NSF and consists of institutional cloud sites deployed at university campuses and labs across the world, connected to research networks capable of providing on-demand virtual connection services (Internet2, NLR, ESnet~\cite{esnet}). In less than two years of limited operation, ExoGENI has served more than 5000 slices to over a hundred unique users across a growing number of geographically distributed sites. ExoGENI testbed is managed by distributed software called ORCA (Open Resource Control Architecture~\cite{Chase:ORCA}) that performs multi-cloud orchestration across these sites. ORCA uses semantic technologies to drive resource orchestration decisions to create user slices and this functionality represents the focus of the paper.

ORCA is a distributed system that has a number of actor types and many instances of each actor type, some of which are associated with individual resource providers, some serve as coordination points for distributed resource scheduling and allocation and some as the entry points for the users to place their slice requests with the system. Actors communicate with each other using internally-defined web services protocols. Importantly, semantic resource descriptions encoded as RDF-XML documents are exchanged between these actors using these internal protocols.

\begin{wrapfigure}{r}{0.5\textwidth}
\vspace{-0.5in}
  \begin{center}
    \includegraphics[width=0.48\textwidth]{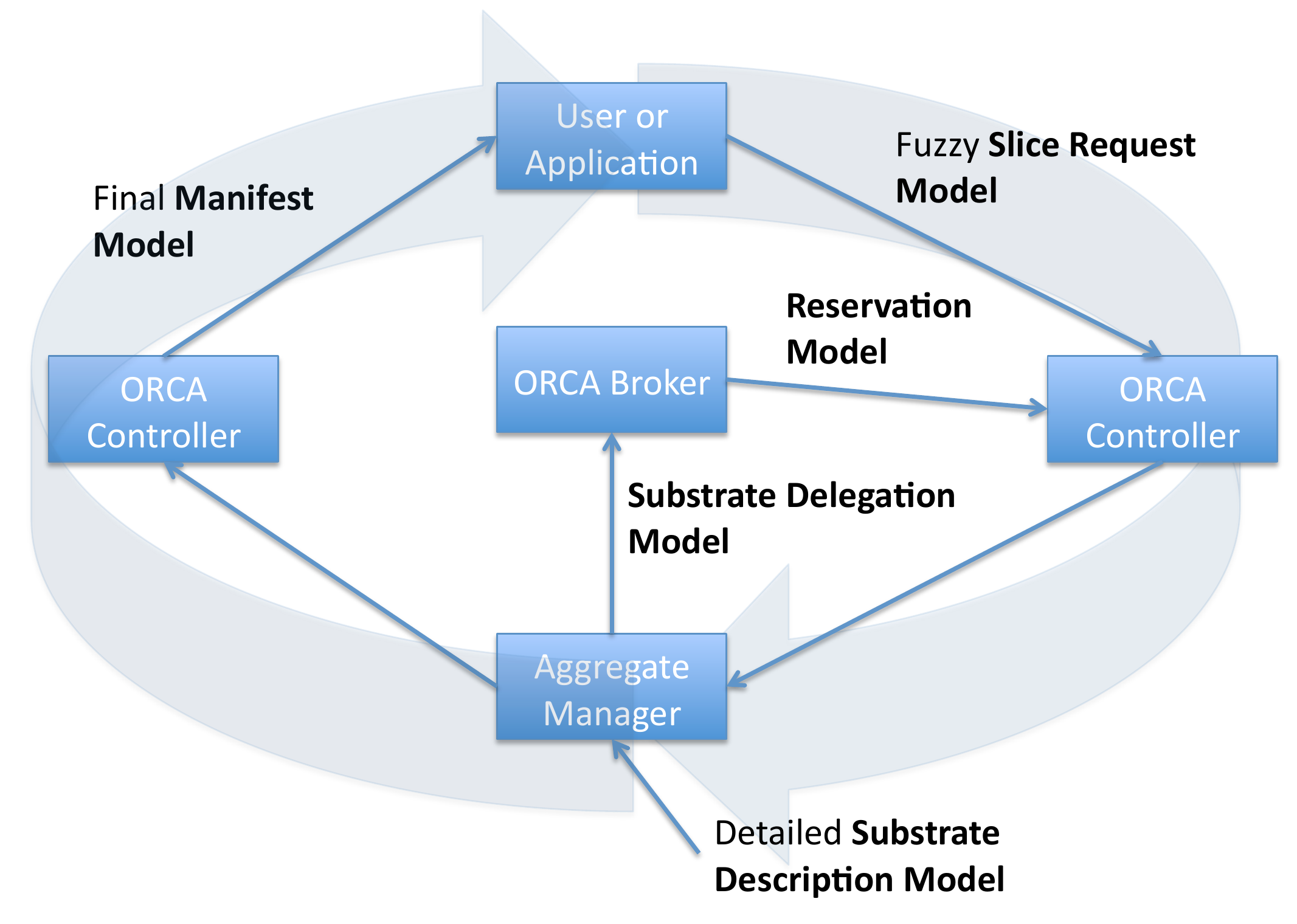}
  \end{center}
  \caption{Semantic resource models in ORCA. }
  \label{fig:cycle}
\vspace{-0.2in}
\end{wrapfigure}

The example flow of these documents is described in Figure~\ref{fig:cycle}. The process begins with the resource provider supplying a detailed description of its resources and their topology to the {\em ORCA Aggregate Manager} or AM - an actor type responsible for representing resource providers. There is typically a one-to-one correspondence between a resource provider and a specific AM. This description, which conforms to the {\em Substrate Description} model, in the form of an RDF-XML file is loaded by the AM and processed by it. The description is transformed into another document, this time conforming to the {\em Substrate Delegation} model, which is passed on to the selected {\em ORCA Broker} - an actor responsible for coordination of resource allocations across providers. The purpose of the Substrate Delegation model is to compress the detailed topological representation of a provider into something that is much smaller and less detailed, protects their privacy and is suitable for inter-domain path-finding.

An ExoGENI user provides the selected {\em ORCA Controller} with an RDF-XML document describing the topology and properties of the desired slice, that conforms to the {\em Slice Request} model. The document is processed by the topology embedding workflow using SPARQL queries, inferences and procedural code. The controller then requests from the Broker the available resources and designs a slice manifest, conforming to the {\em Slice Manifest} model, which describes the details of the slice 'as-built'. The manifest topology is an iso- or homeomorphic mapping of the request onto the graph describing the topology of the providers. The manifest contains information about which specific resources were instantiated and any details needed by the user to operate the resources. 

Based on the information in the manifest, the controller communicates with individual AMs to provision and interconnect elements of the slice, in parallel filling out the details of the manifest model. Importantly, AMs update their internal semantic models reflecting the current use of their resources and the controllers update their global views of known used resources in various aggregates by inserting new facts into the models, like provisioned hosts or network paths. Finally, when the slice is ready, the manifest is returned to the user, again, as an RDF-XML document. 


The ExoGENI ontologies consist of two parts: 

{\bf The static class and property vocabularies} hosted at \url{http://geni-orca.renci.org/owl}. These are OWL schemas, mostly T-boxes, with a few A-boxes related to permanent elements of the infrastructure. They consist of on the order of 6500 statements with approximately 1500 classes and several hundred object and data properties, which have been validated for consistency using Pellet (v1.5) reasoning engine built into Proteg{\'e}. 

{\bf The declarative resource descriptions} exchanged by the ORCA actors, consisting exclusively of A-boxes - assertions about the state and relationships between network resources, that reference the T-boxes in the static ontologies. The number of statements in these is linear with the number of hardware resource elements being described. These are constructed either a-priori, as in the case of user slice requests to controllers or detailed substrate descriptions supplied by resource providers to AMs, or on the fly, as is the case with slice manifests constructed and supplied by controllers to users.
At processing time, the ontologies with A-boxes are merged with ontologies with T-boxes into a single OWL DL model, which enables inferences and thus more capable queries on the resources described within the documents. 

As a final note on the implementation, ORCA actors utilize the Jena library for creating, manipulating and querying the various models. We use internal Jena inference engine for running entailments and the built-in Datalog-like rule engine for rule-based inferences.

{\bf ExoGENI ontologies}

\begin{figure}[t]
        \centering
        \begin{subfigure}[b]{0.8\textwidth}
                \centering
                \includegraphics[width=\textwidth]{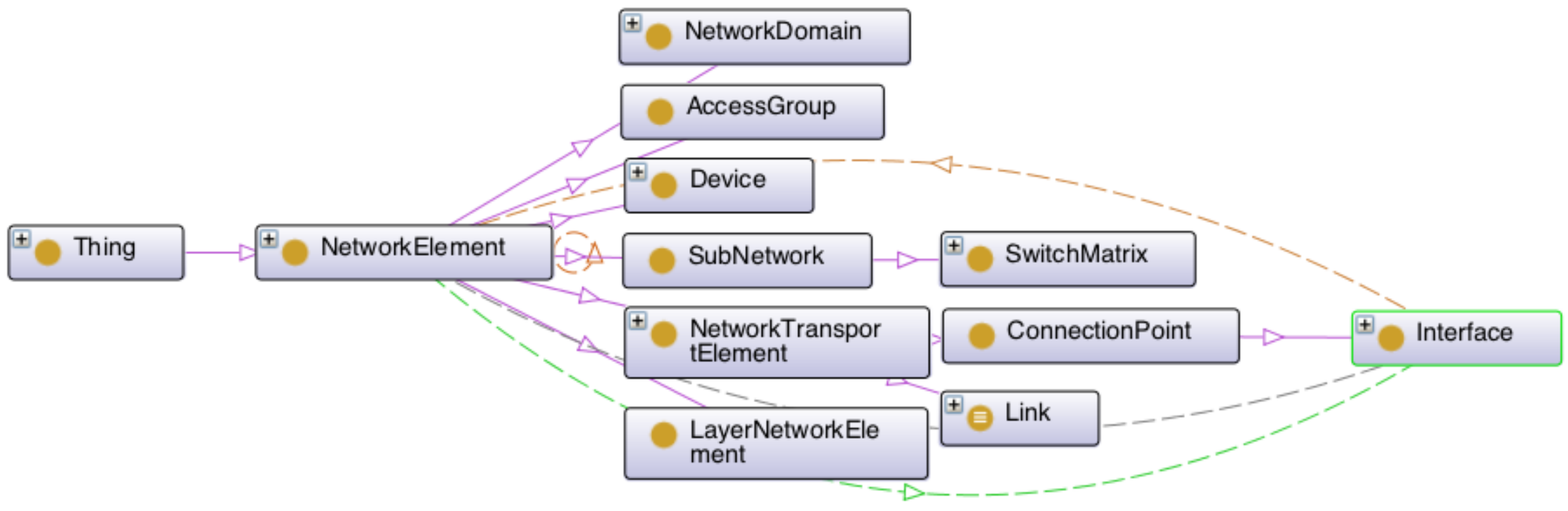}
                \label{fig:topo}
        \end{subfigure}%
        ~ 
        \begin{subfigure}[b]{0.2\textwidth}
                \centering
                \includegraphics[width=\textwidth]{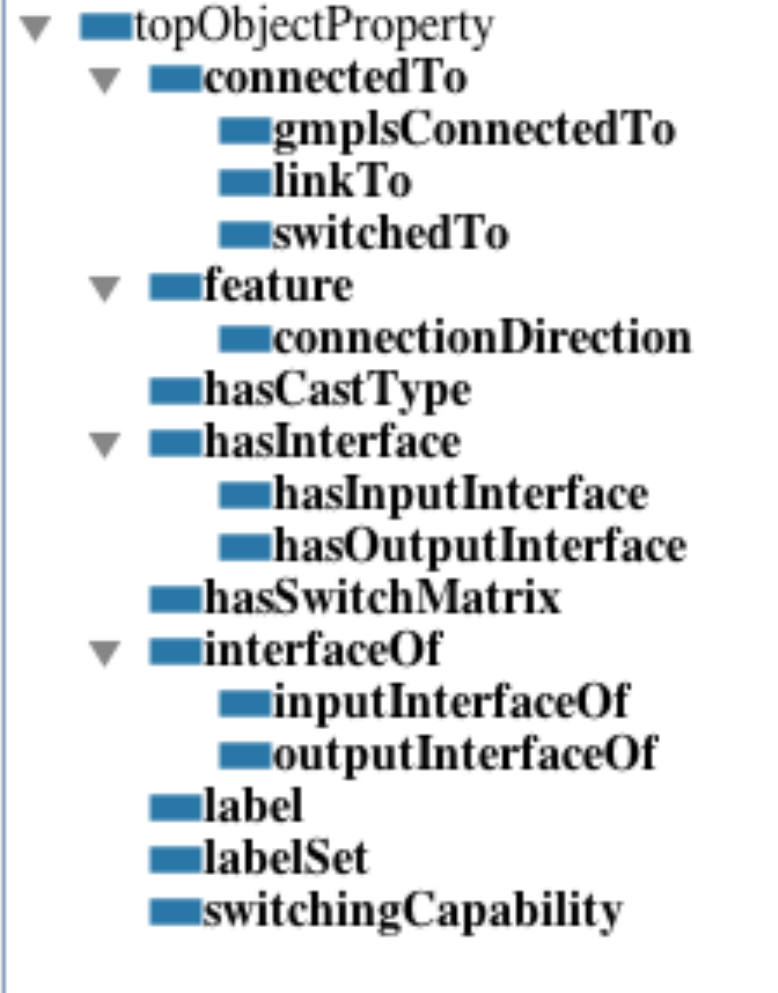}
                \label{fig:props}
        \end{subfigure}
        \caption{Basic classes and object properties of top-level topology ontology}
        \label{fig:topo-all}
\end{figure}

Our ontology, called NDL-OWL, provides a vocabulary to describe elements of compute, storage and network infrastructure and how they are interconnected with each other to aid in path finding, topology embedding and other types of resource management activities. The ultimate goal of this  process is to create a representation language that is {\em sufficiently powerful to enable generic resource control modules to reason about networked resources and the ways that the system might share them, partition them, and combine them}. The top-level ontology for this is used for describing the high level abstraction of a network topology - topology.owl (see Figure~\ref{fig:topo-all}). 

This schema defines a hierarchy of basic classes and object properties needed to describe network topology abstractions: everything begins with a base class called {\em NetworkElement} that represents any possible resource within a network. Subclassed off it are {\em NetworkDomain}s, which represent groupings of resources under a single administrative control, {\em Device}s, which represent end-points, {\em NetworkTransportElement}s, a subclass dedicated to elements of the network through which bits transit - i.e. interfaces and links of various types and so on. The object properties help relate various  network elements to each other i.e. {\em connectedTo}, {\em hasInterface} and its inverse property {\em interfaceOf} that associates nodes, links and their interfaces and {\em label} which allows to associate a variety of label types with network elements e.g. IP addresses, VLAN or MPLS tags, Ethernet MAC addresses and so on. This latter property is critical to properly interconnecting elements of the infrastructure with each other as labels must be negotiated to allow connections e.g. a compute node must be told which VLAN tag to attach itself to in order to properly connect to other nodes.

\begin{figure}[t]\begin{center}  
\includegraphics[width=0.7\textwidth]{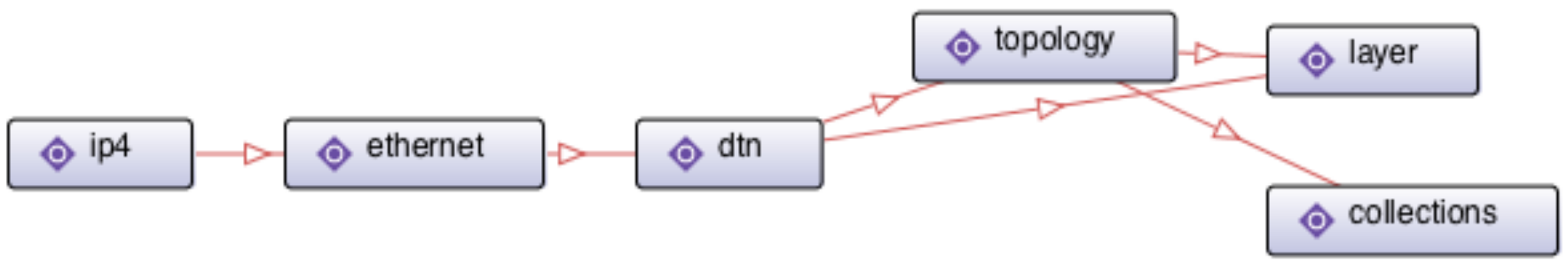}        
	\caption{Ontology import graph for multi-layered networks.}
	\label{fig:imports}
\end{center}
\vspace{-0.3in}
\end{figure}

NDL-OWL defines subordinate ontologies that help define multiple layers of transport and routed networks (consistent with~\cite{ITU:G805,ITU:G809}) - e.g. optical (dtn.owl) which describe connectivity in terms of optical wavelengths and timeslots within those, Layer 2 (ethernet.owl), which describe connectivity in terms of VLANs, IP (ip4.owl), which provides features necessary to describe an IPv4 network (IP addresses, netmasks as labels etc) shown in Figure~\ref{fig:imports}. 

\begin{figure}[t]
        \centering
        \begin{subfigure}[b]{0.74\textwidth}
                \centering
                \includegraphics[width=\textwidth]{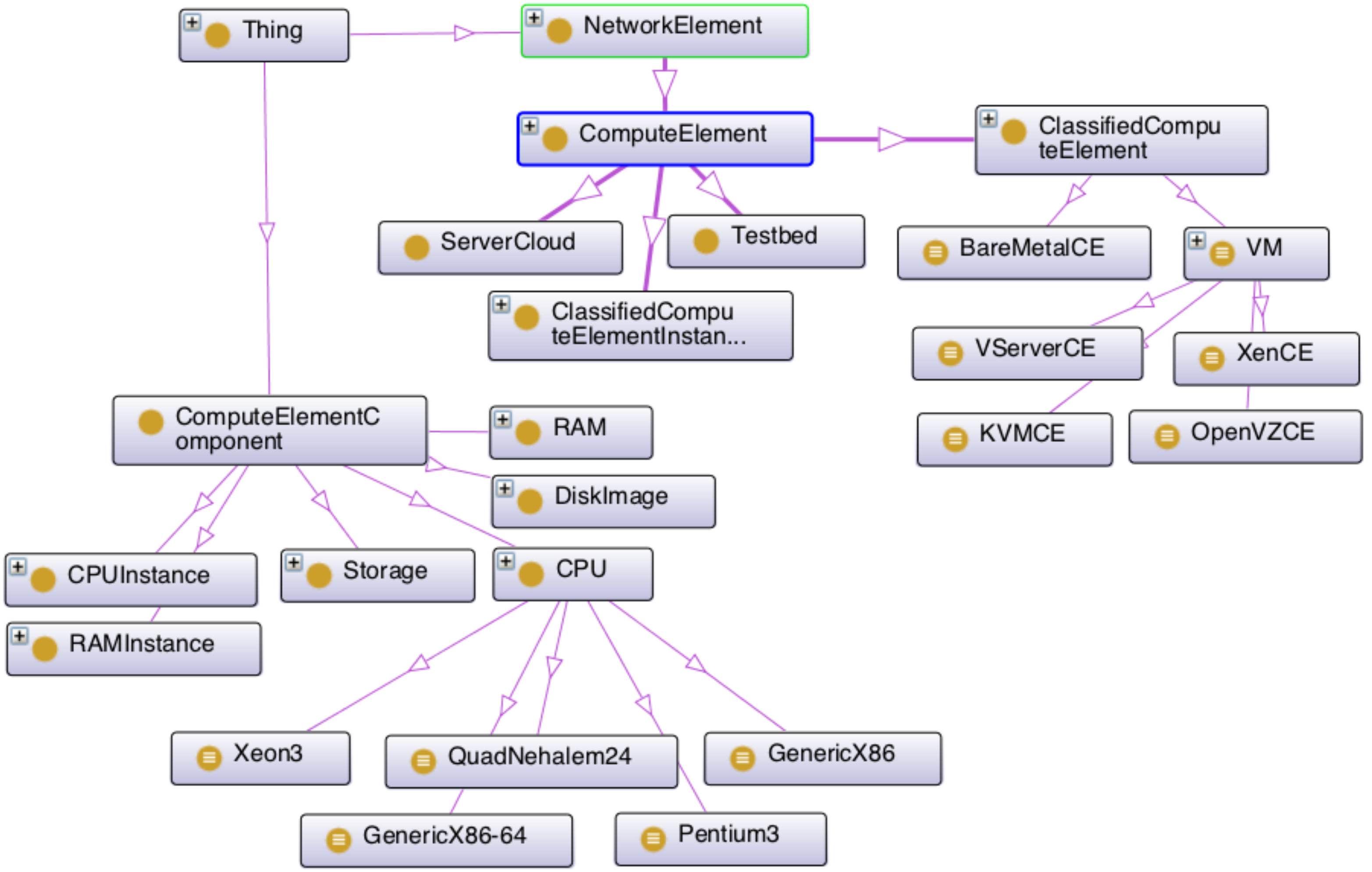}
                \label{fig:topo}
        \end{subfigure}%
        ~ 
        \begin{subfigure}[b]{0.26\textwidth}
                \centering
                \includegraphics[width=\textwidth]{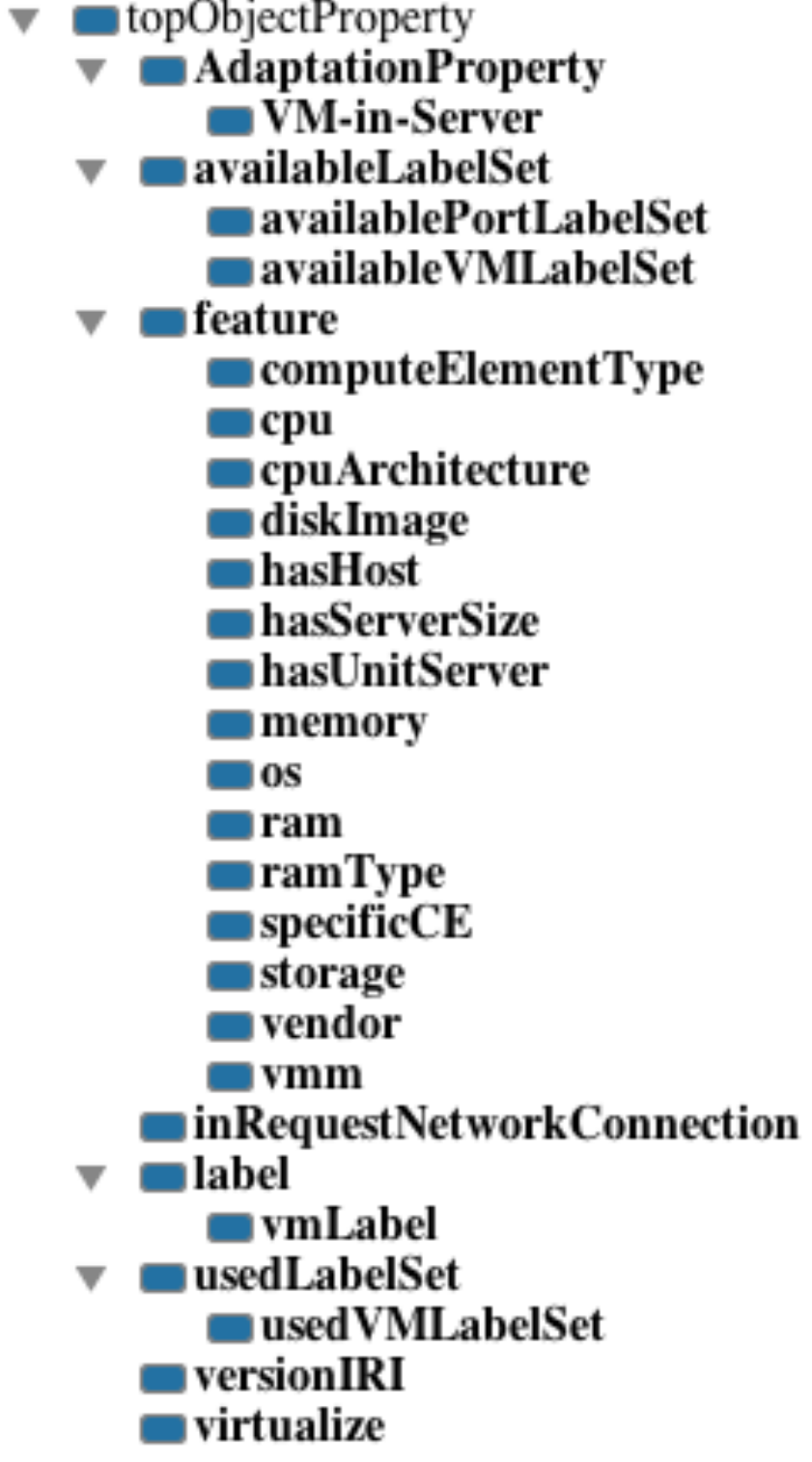}
                \label{fig:props}
        \end{subfigure}
        \caption{Basic classes and object properties of compute ontology}
        \label{fig:compute-all}
\end{figure}

We added a number of ontologies required to describe the edge compute and storage resources. Figure~\ref{fig:compute-all} shows the class hierarchy of the top-level compute ontology, which starts with a  {\em ComputeElement} class as a subclass of {\em NetworkElement} from the upper topology ontology. A ComputeElement is further subclassed into {\em ServerCloud}s, {\em Testbed}s and {\em ClassifiedComputeElement}s. The first two are ways of abstracting multiple physical compute elements into a simplified definition, used e.g. for delegating resources in the Delegation Model described above. The ClassifiedComputeElements is a subtree of classes describing various types of compute elements available in ExoGENI - {\em BareMetalCE}s - compute elements that are 'bare-metal', i.e. provisioned directly without any virtualization and {\em VM}s - compute elements provisioned using a variety of virtualization techniques (VServer and OpenVZ containers, KVM and Xen hypervisors~\cite{xen}). The details of these are not crucial for this paper, however it is important to note that different types of virtualization offer different performance isolation properties and are used by different providers in ExoGENI testbed, therefore it is important to enable users to request compute nodes with different types of virtualization. 



Importantly, none of these ontologies need the vocabulary to describe the inner workings of each infrastructure element, e.g. a router or a compute node. Instead they must provide enough information about {\em features and connectivity between them} to support the resource selection and topology embedding tasks, common in the NIaaS environment.

{\bf Using NDL-OWL to describe NIaaS resources}

\begin{figure}[t]\begin{center}  
\includegraphics[width=0.8\textwidth]{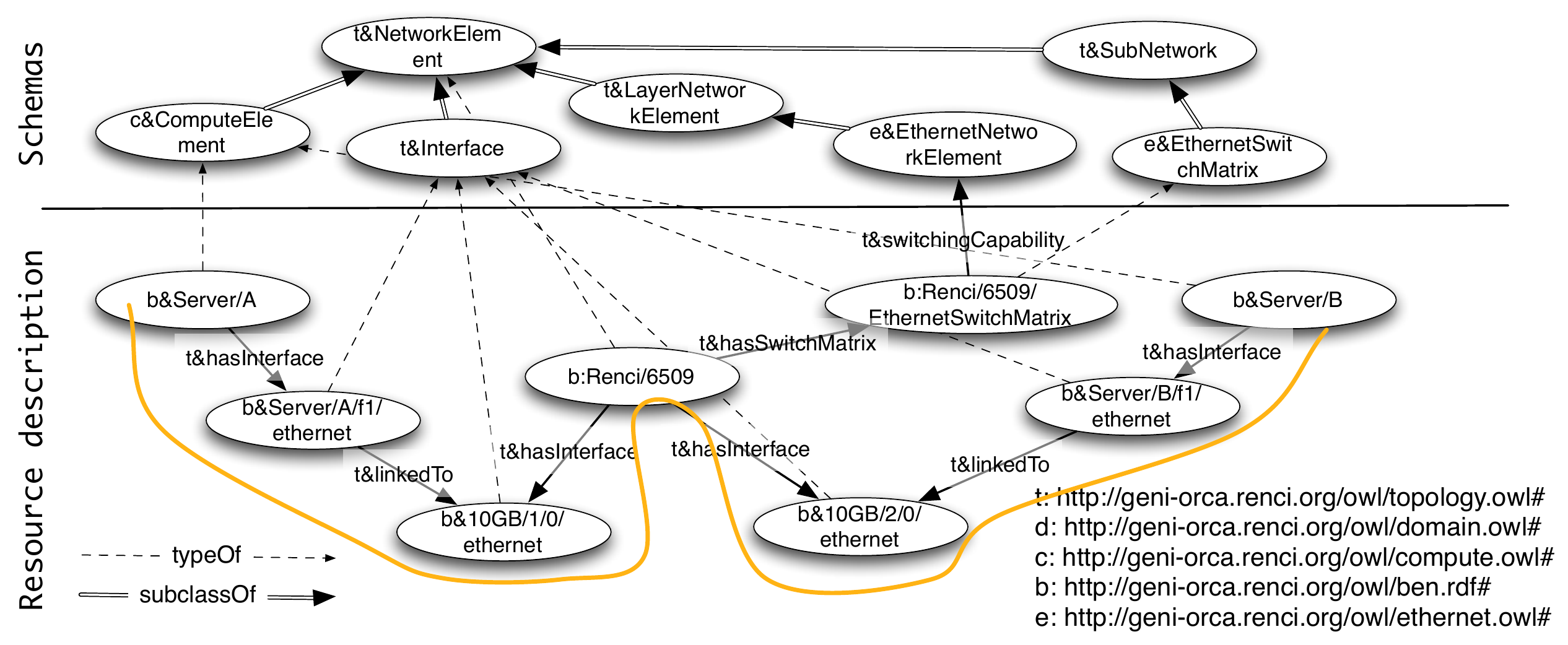}        
	\caption{Simple advertisement of connection between two servers.}
	\label{fig:ad}
\end{center}
\vspace{-0.4in}
\end{figure}

This section demonstrates how the ontologies we described are used to advertise available resources, request resources from the system or support the provisioning actions taken by the system.

Figure~\ref{fig:ad} shows a very simple {\em advertisement} by a provider of two hardware servers (Server/A and Server/B) connected by an Ethernet switch (Renci/6509). The diagram does not show most of the compute element details, concentrating instead on the means to describe topology and connectivity. The curved yellow line shows the network topology as a subgraph embedded into the semantic graph annotated with other necessary information. A switch matrix of type {\em EthernetNetworkElement} indicates that the connection is at Ethernet and not any other layer, which indicates a path constraint. The connection between a server and a switch can be extracted by following the {\em hasInterface} property from Server/A to Server/A/f1/ethernet, then via {\em linkedTo} property, indicating a presence of either a physical or virtual link, to an interface 10GB/1/0/ethernet that belongs to the Renci/6509 switch. 

\begin{figure}[t]\begin{center}  
\includegraphics[width=0.8\textwidth]{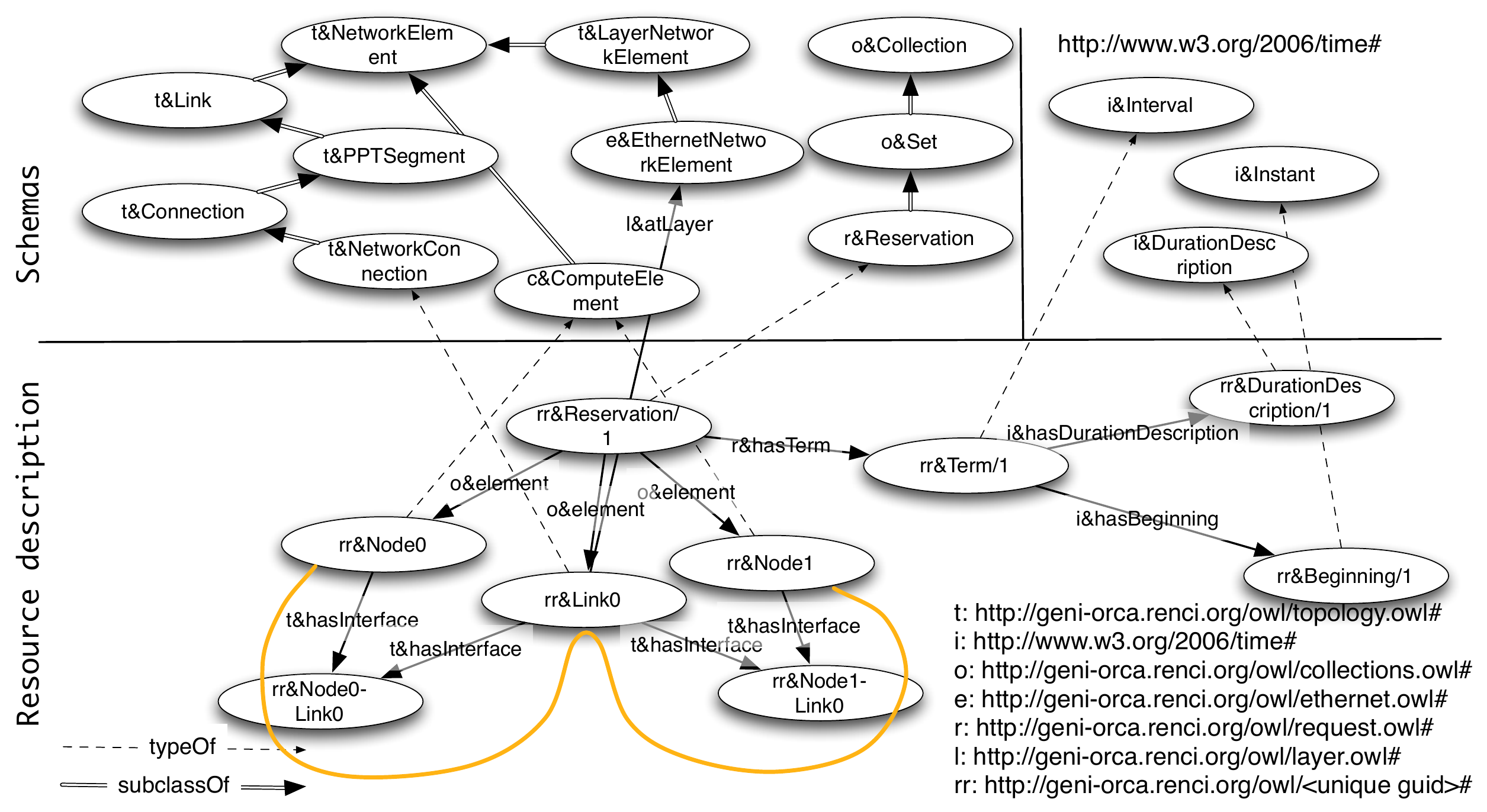}        
	\caption{Simple request for a slice with two nodes connected by a link.}
	\label{fig:request}
\end{center}
\vspace{-0.4in}
\end{figure}

A more complex example in Figure~\ref{fig:request} shows a {\em request} by a user for a topology that has two nodes and a link. The goal of the system is to embed this request in available substrate by finding a homeomorphic mapping. Again, the curved line shows the actual requested network topology embedded into the semantic graph. In addition to nodes that are of type {\em ComputeElement} and a link of type {\em NetworkConnection} within Ethernet layer indicated by {\em atLayer} property pointing to {\em EthernetNetworkElement}, this semantic model also has other entities that make it conform to Request model. Namely this is the {\em Reservation/1} entity of type {\em Reservation} from a request.owl ontology we have defined, which acts as a collection of requested elements, by using {\em element} property to point to requested nodes and links. It also points to the desired start time and duration of this request by pointing to {\em Term/1} entity of type {\em Interval} that has a beginning and a duration. The classes related to describing time intervals come from a well-known time ontology vocabulary http://www.w3.org/time\#. 

The manifest model (not shown) describes the topology and meta information of the provisioned slice in a similar fashion. It also includes all the statements from the slice request model and has specialized object properties linking entities of the provisioned resources to the entities of the request, to indicate exactly which element in the request corresponds to which element of the provisioned infrastructure. This feature is essential for automated processing of the manifests by other systems, which need to request and operate on provisioned slices.

{\bf Using Semantic Queries}

In order to produce a manifest of the slice, the system must find an embedding of the request topology graph in the graph describing the interconnected topology of the various providers. This embedding must satisfy a number of constraints, including resource availability, path continuity, layering, adaptations as well as bandwidth and latency.

The network path embedding problem has two  parts: (1) finding one or more {\em feasible} constrained shortest path in the network topology consisting of multiple domains and (2) finding the internal elements of the path, i.e., all the devices, layers, and interfaces so that {\em configuration commands can be correctly formed} to each network element to actually provision the connections on the path.   

Network path computations are similar to query subgraph extraction in the corresponding RDF graph. For example, a path in the network topology is equivalent to a property path in the semantic graph. Complex pattern queries using SPARQL, can be computationally intractable as studied in~\cite{perez2006semantics}.  The query evaluation can be done in polynomial time
if the pattern only contains the AND and Filter operators. The evaluation becomes NP-Complete if AND, FILTER, and
UNION operators appear in the pattern. If OPT operator is involved, the problem becomes a PSPACE-complete problem. Incidentally, computing bandwidth constrained shortest path in networks is NP-Complete~\cite{bandwidth-delay-np}.

To implement an inter-domain bandwidth-constrained path-finding algorithm, ORCA relies on a heuristic that combines Gleen-enhanced SPARQL queries with Dijkstra's shortest path algorithm. Gleen is a regular path expression library plugin to the Jena ARQ package~\cite{detwiler2008regular}. The output of the algorithm is used to perform provisioning of resources and embedding of customer slice topologies into the topology of multiple providers. Gleen supports the regular expression operators like '?' (zero or one), '*' (zero or more), '+' (one or more), '|' (alternation), and '/' (concatenation). Gleen was designed to find path patterns between two entities in a medical ontology so that a simplified view can be generated out of the complicated class and property hierarchies.  It defines two types of query constructs that can be directly applied to a triple pattern of the SPARQL query body. In both cases, the path expression is formed by a number of properties recursively using the regular expression language.

SPARQL 1.1 offers support for regular path expressions, with some limitations~\cite{arenas2012counting}.  However it does not offer a way to specify the path constraints that the orchestration process needs, and there are no query constructs to return the internal path elements through a path query - the elements crucial to forming the provisioning commands on networking equipment. 

The {\it gleen:OnPath} construct is used to find all the objects that are connected to the subject via the defined path expression. A triple pattern of this construct can be expressed as:  

{\it \small
\noindent subject gleen:OnPath (pathExpression object)
}


\begin{figure}[t]
	\centering
	\begin{subfigure}[b]{0.48\textwidth}
		\centering
		\includegraphics[width=\textwidth]{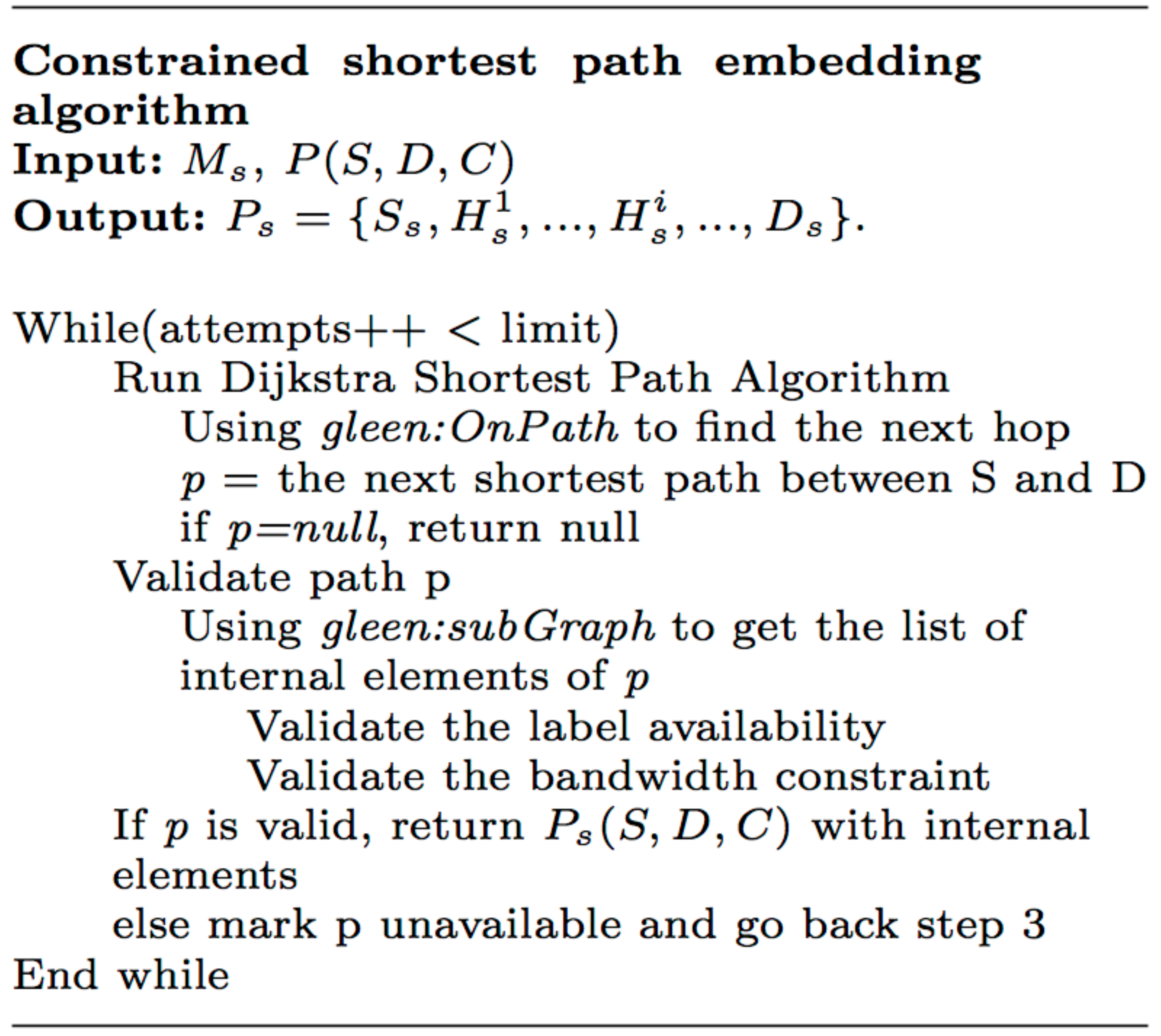}
		\label{fig:algo}
	\end{subfigure}
	\begin{subfigure}[b]{0.48\textwidth}
		\centering
		\includegraphics[width=\textwidth]{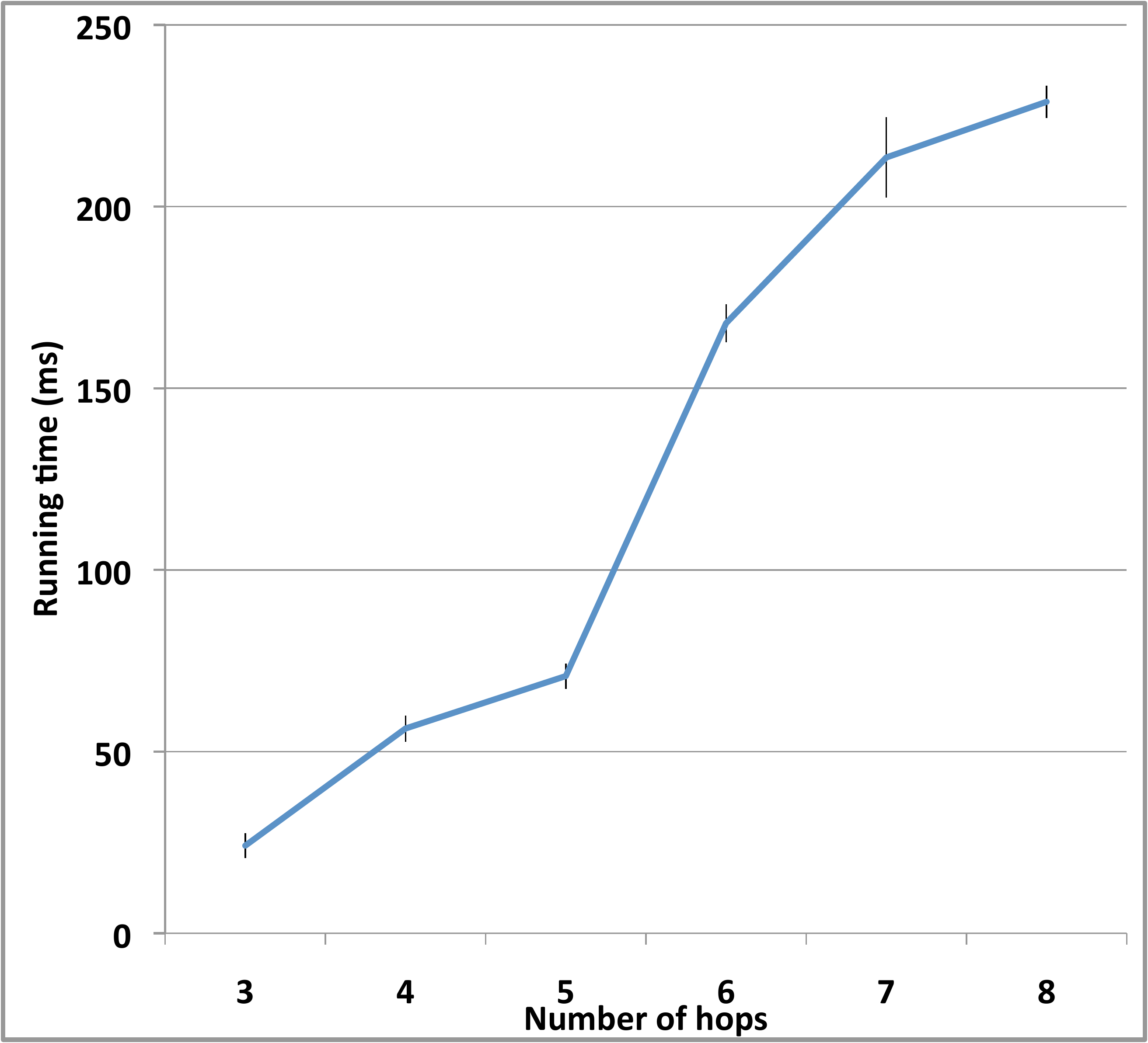}
		\label{fig:perf}
	\end{subfigure}
	\caption{Multi-domain constrained shortest path embedding algorithm and its performance (MacBook Pro 2.4GHz Intel Core 2 Duo, 4GB RAM, Java 1.6.0\_43, Jena 2.6.0; 20 runs).}
	\label{fig:algo-all}
	\vspace{-0.2in}
\end{figure}

The following simple query returns all the network devices that are reachable from a specific {\it source} device via one or more hops.

{\it \small
\begin{tabbing}
\= \ \ \ \ \ \ \= \ \ \ \ \= \ \ \ \ \= \ \ \ \ \= \ \ \ \ \=\kill
\>Select ?destination 	\\
\>Where \{   \\
\>\>source gleen:OnPath  \\
\>\> (([ndl:hasInterface]+/[ndl:connectedTo]+/[ndl:interfaceOf]+)+ ?destination ). \\
\>\} \\
\end{tabbing}					
}
\vspace{-0.3in}

However, this pattern only returns the destination objects without revealing how the paths are traversed. The second construct {\it gleen:Subgraph} is defined to accomplish this and can be applied to the SPARQL triple pattern in the following way:

{\it \small
\noindent (inputSubject pathExpression inputObject) gleen:Subgraph (outputSubject outputPredicate outputObject) .
}

The three arguments in the object position triple must be unbound and are the variables to be answered by the query. In this way, all intermediate resources along with the path edges connecting them are obtained for path between {\it inputSubject} and {\it outputSubject} via the {\it pathExpression} found via the first query.

We use the following query to find every hop that a path between device {\it source} and {\it destination} traverses within the same network layer.

{\it \small
\begin{tabbing}
\= \ \ \ \ \ \ \= \ \ \ \ \= \ \ \ \ \= \ \ \ \ \= \ \ \ \ \=\kill
\>Select ?a ?b ?c 	\\
\>Where \{   \\
\>\>(source   \\
\>\> ([ndl:hasInterface]+/[ndl:connectedTo]+/[ndl:interfaceOf]+)+ destination). \\
\>\>gleen:Subgraph (?a ?b ?c) \\
\>\} \\
\end{tabbing}					
}
\vspace{-0.3in}

Our path-finding heuristic is depicted in Fig.~\ref{fig:algo-all}. It is similar to a K edge-disjoint shortest path algorithm~\cite{bhandari}, however instead of generating $k$ candidate paths upfront it generates them as it goes marking traversed paths as unavailable. This approach has known limitations, however works well under our conditions~\cite{bhandari}. Our algorithm takes the substrate RDF model $M_s$ and a path request $P(S,D,C)$ as the input, where $S$, $D$, and $C$ are the source, destination points, and the path constraints respectively. It uses Gleen queries to construct partial solutions that are checked for validity.
It returns a list of network elements the shortest path that satisfy the constraints if there is one. 

Figure~\ref{fig:algo-all} shows the performance of the algorithm on a real multi-domain topology in ExoGENI, where paths of increasing number of hops must be computed with enough detail to be provisioned (the provisioning time is highly variable and is not included in the graph). Due to the sparsity of the multi-domain graph, which is a function of ExoGENI existing deployments (a total of 31 provider domains), the algorithm exhibits pseudo-linear behavior. As the degree of connectivity of the multi-domain graph increases, the running time of the algorithm, in which the Gleen queries dominate, would also be expected to increase. 

%
%
%

{\bf Inferences}

ORCA uses OWL entailments to simplify its topology embedding algorithms. While, for example, there can be many types of nodes, i.e. different compute nodes (virtual machines, bare-metal nodes of different types) or network switches and there can be different types of links, what matters to the pathfinding algorithm is what layer the requested connection is at, the layers of available connection segments, the adaptation capabilities available within nodes on the path to cross layers and which domains the end nodes belong to. ORCA runs standard RDFS and OWL entailments, which enables the algorithm to use these generic concepts in its SPARQL/Gleen queries, rather than operating on exact resource types. Once the path is computed, including its intermediate elements, the exact resource types are used for determining the provisioning actions that need to take place, which are specialized to each provider and are separate from the pathfinding algorithm. 

This approach keeps path finding and topology embedding algorithms generic, allowing ExoGENI to easily incorporate new resource types as it evolves and incorporates new resource provider domains. 

Another use for inferences in ExoGENI is the validation of slice topology requests from the users provided to ORCA. In our environment a request, in the form of RDF-XML document, can be produced by a number of entities and tools of unknown origin. Prior to processing the request, ORCA must ensure that semantically it makes sense. While schema validation performs some of the necessary checks, there are limitations to the expressivity of the schema, which require augmenting this process. Procedural verification is not portable and hard to ensure correctness and consistency across implementations. Instead we use a set of semantic rules expressed as a subset of Datalog (only arity one and two predicates are allowed), to describe these additional constraints, which are executed by Jena Datalog engine.

For example, if a user is attempting to embed a broadcast connection (one with more than two endpoints) that connects multiple domains, each domain must be mentioned only once. E.g. it is OK to say, `I would like to have a broadcast connection between nodes belonging to domains A, B and C'. It is NOT OK to say `I would like to have a broadcast connection between nodes belonging to domains A, B and A', since this actually represents a poorly formed request for a point-to-point connection. The user tool must re-normalize the request prior to submitting. The rule expressing this constraint is shown below:


\vspace{-0.1in}
{\it \small
\begin{tabbing}
\= \ \ \ \ \ \ \= \ \ \ \ \= \ \ \ \ \= \ \ \ \ \= \ \ \ \ \=\kill
(?Z rb:violation error("Domains in broadcast link can't be repeated", ?X))\\
\>$<-$ (?X rdf:type topo:BroadcastConnection), (?X topo:hasInterface ?I1), \\
\>\>(?X topo:hasInterface ?I2), notEqual(?I1, ?I2), (?A topo:hasInterface ?I1), \\
\>\>(?B topo:hasInterface ?I2), (?A rdf:type comp:ComputeElement), \\
\>\>(?B rdf:type comp:ComputeElement), notEqual(?A, ?B), \\
\>\>(?A req:inDomain ?D1), (?B req:inDomain ?D2), equal(?D1, ?D2), \\
\>\>(?X topo:hasInterface ?I3), notEqual(?I1, ?I3), notEqual(?I2, ?I3), \\
\>\>(?C topo:hasInterface ?I3), (?C rdf:type comp:ComputeElement), \\
\>\>(?C req:inDomain ?D3), notEqual(?D3, ?D1)\\
\end{tabbing}					
}
\vspace{-0.2in}

\noindent The set of the rules we use covers other constraints and continues to evolve with the schema and the algorithms.

\vspace{-0.1in}
\section{Conclusions and Future Work}
\label{sec:conclusions}
In this paper we presented an overview of implementation and use of OWL DL-based resource representation models in a multi-domain NIaaS ExoGENI testbed. Using our approach we showed that it is possible to construct an extensible NIaaS system which can (a) use OWL class and role hierarchies to describe the system resources; that (b) the topology embedding algorithms can operate in a generic fashion using a number of standard abstractions built on SPARQL/Gleen queries as building blocks and (c) that models constructed on the fly and exchanged by various software agents can be verified by a combination of standard entailments augmented with portable logic rules that account for the semantics not captured in the OWL schemas, avoiding procedural code. Our work demonstrates the practical uses and future potential of this type of knowledge representation approaches for active management of cyber-resources in a distributed environment on a global scale.

Our ongoing work concentrates on supporting new resource types and providers, increasing the complexity of the topology embedding algorithms and designing an upper ontology for NIaaS testbeds jointly with our colleagues from similarly scoped projects around the world.




 

\vspace{-0.1in}
\vspace{-0.1in}


\bibliographystyle{abbrv}
\bibliography{refs,bib,vnet,sdci,clouds,duke,pegasus,optics,sdci-duke,IRNC09,semantic}

\begin{thebibliography}{10}

\bibitem{i2:dcn}
{Internet2 Dynamic Circuit Network (DCN)}.
\newblock \url{http://www.internet2.edu/network/dc/}.

\bibitem{NDL:GLIF}
{NDL} demonstration site at {GLIF}.
\newblock http://ndl.uva.netherlight.nl/.

\bibitem{NDL:main}
Network {D}escription {L}anguage.
\newblock http://sne.science.uva.nl/ndl/.

\bibitem{ec2}
{Amazon.com, Inc.}
\newblock {Amazon Elastic Compute Cloud (Amazon EC2)}.
\newblock \url{http://www.amazon.com/ec2}.

\bibitem{arenas2012counting}
M.~Arenas, S.~Conca, and J.~P{\'e}rez.
\newblock {Counting beyond a Yottabyte, or how SPARQL 1.1 property paths will
  prevent adoption of the standard}.
\newblock In {\em Proceedings of the 21st international conference on World
  Wide Web}, pages 629--638. ACM, 2012.

\bibitem{exogeni-tridentcom12}
I.~Baldine, Y.~Xin, A.~Mandal, P.~Ruth, A.~Yumerefendi, and J.~Chase.
\newblock {ExoGENI: A Multi-Domain Infrastructure-as-a-Service Testbed}.
\newblock In {\em TridentCom: International Conference on Testbeds and Research
  Infrastructures for the Development of Networks and Communities}, June 2012.

\bibitem{xen}
P.~Barham, B.~Dragovic, K.~Fraser, S.~Hand, T.~Harris, A.~Ho, R.~Neugebauer,
  I.~Pratt, and A.~Warfield.
\newblock {Xen and the Art of Virtualization}.
\newblock In {\em Proceedings of the 19th ACM Symposium on Operating Systems
  Principles (SOSP)}, pages 164--177, Bolton Landing, NY, October 2003.

\bibitem{bhandari}
R.~Bhandari.
\newblock {\em {Survivable Networks: Algorithms for diverse routing}}.
\newblock Kluwer Academic Publishers, 1999.

\bibitem{Chase:ORCA}
J.~Chase, L.Grit, D.Irwin, V.Marupadi, P.Shivam, and A.Yumerefendi.
\newblock Beyond virtual data centers: Toward an open resource control
  architecture.
\newblock In {\em Selected Papers from the International Conference on the
  Virtual Computing Initiative (ACM Digital Library)}, May 2007.

\bibitem{Corcho2006a}
O.~Corcho, P.~Alper, I.~Kotsiopoulos, P.~Missier, S.~Bechhofer, and C.~Goble.
\newblock {An overview of S-OGSA : A Reference Semantic Grid Architecture}.
\newblock 4:102--115, 2006.

\bibitem{detwiler2008regular}
L.~T. Detwiler, D.~Suciu, and J.~F. Brinkley.
\newblock {Regular paths in SPARQL: Querying the NCI thesaurus}.
\newblock In {\em AMIA Annual Symposium Proceedings}, volume 2008, page 161.
  American Medical Informatics Association, 2008.

\bibitem{ExoGENI:www}
{ExoGENI website and wiki}.
\newblock \url{http://www.exogeni.net, http://wiki.exogeni.net}.

\bibitem{rspec}
GENI.
\newblock {GENI RSpec v3}.
\newblock http://groups.geni.net/geni/wiki/RSpecSchema3.

\bibitem{esnet}
C.~Guok, D.~Robertson, E.~Chaniotakis, M.~Thompson, W.~Johnston, and
  B.~Tierney.
\newblock {A User Driven Dynamic Circuit Network Implementation}.
\newblock In {\em DANMS, IEEE}, 2008.

\bibitem{Haase}
P.~Haase, T.~Math, M.~Schmidt, A.~Eberhart, and U.~Walther.
\newblock {Semantic Technologies for Enterprise Cloud Management}.
\newblock {\em Proceedings of the 9th international semantic web conference}.

\bibitem{NDL:Overview}
J.~Ham, F.~Dijkstra, P.~Grosso, R.~Pol, A.~Toonk, and C.~Laat.
\newblock {A Distributed Topology Information System for Optical Networks Based
  on the Semantic Web}.
\newblock {\em Journal of Optical Switching and Networking}, 5(2-3), June 2008.

\bibitem{PerfSonar}
A.~Hanemann, J.~W. Boote, E.~L. Boyd, J.~Durand, L.~Kudarimoti, R.~Lapacz,
  D.~M. Swany, J.~Zurawski, and S.~Trocha.
\newblock {PerfSONAR: A Service Oriented Architecture for Multi-Domain Network
  Monitoring}.
\newblock In {\em Proceedings of the Third International Conference on Service
  Oriented Computing}, volume LNCS 3826, pages 241--254. Springer-Verlag, 2005.

\bibitem{ITU:G805}
ITU-T.
\newblock G.805 : Generic functional architecture of transport networks.

\bibitem{ITU:G809}
ITU-T.
\newblock G.809: Functional architecture of connectionless layer network,
  http://www.itu.int/rec/t-rec-g.809.

\bibitem{nist-cloud}
P.~Mell and T.~Grance.
\newblock {The NIST Definition of Cloud Computing}.
\newblock {Special Publication 800-145, Recommendations of the National
  Institute of Standards and Technology}, September 2011.

\bibitem{Moscato2011}
F.~Moscato, B.~D. Martino, and V.~Munteanu.
\newblock {An Analysis of mOSAIC ontology for Cloud Resources annotation}.
\newblock {\em Computer Science and Information Systems (FedCSIS), 2011
  Federated Conference on}, pages 973--980, 2011.

\bibitem{eucalyptus}
D.~Nurmi, R.~Wolski, C.~Grzegorczyk, G.~Obertelli, S.~Soman, L.~Youseff, and
  D.~Zagorodnov.
\newblock {The Eucalyptus Open-Source Cloud-Computing System}.
\newblock In {\em {Proceedings of the 9th IEEE/ACM International Symposium on
  Cluster Computing and the Grid (CCGRID)}}, May 2009.

\bibitem{Nyren2011}
R.~Nyren, A.~Edmonds, A.~Papaspyrou, and T.~Metsch.
\newblock {Open Cloud Computing Interface - Core, GFD-P-R.183, OCCI-WG}.
\newblock Technical report, 2011.

\bibitem{OpenStack:www}
{OpenStack Cloud Software}.
\newblock \url{http://openstack.org}.

\bibitem{parsia2004pellet}
B.~Parsia and E.~Sirin.
\newblock {Pellet: An OWL DL reasoner}.
\newblock In {\em Third International Semantic Web Conference-Poster}, page~18,
  2004.

\bibitem{perez2006semantics}
J.~P{\'e}rez, M.~Arenas, and C.~Gutierrez.
\newblock {Semantics and Complexity of SPARQL}.
\newblock In {\em The Semantic Web-ISWC 2006}, pages 30--43. Springer, 2006.

\bibitem{shearer2008hermit}
R.~Shearer, B.~Motik, and I.~Horrocks.
\newblock {HermiT: A highly-efficient OWL reasoner}.
\newblock In {\em Proceedings of the 5th International Workshop on OWL:
  Experiences and Directions (OWLED 2008)}, pages 26--27, 2008.

\bibitem{Tahamtan2012}
A.~Tahamtan, S.~A. Beheshti, A.~Anjomshoaa, and a.~M. Tjoa.
\newblock {A Cloud Repository and Discovery Framework Based on a Unified
  Business and Cloud Service Ontology}.
\newblock {\em 2012 IEEE Eighth World Congress on Services}, pages 203--210,
  June 2012.

\bibitem{NDL:optical-nets}
J.~van~der Ham, P.~Grosso, R.~van~der Pol, A.~Toonk, and C.~de~Laat.
\newblock Using the network description language in optical networks.
\newblock In {\em Tenth IFIP/IEEE Symposium on Integrated Network Management},
  May 2007.

\bibitem{bandwidth-delay-np}
Z.~Wang and J.~Crowcroft.
\newblock {Bandwidth-delay based routing algorithms}.
\newblock In {\em Global Telecommunications Conference, 1995. GLOBECOM '95.,
  IEEE}, volume~3, pages 2129--2133 vol.3, 1995.

\end{thebibliography}


\end{document}